\documentclass[aps,twocolumn,amsmath,amssymb,showpacs,prb,superscriptaddress,unsortedaddress]{revtex4}
\usepackage{epsf}
\usepackage{graphicx}

\begin{document}

\title{Direct observation of a Fermi surface and superconducting gap in LuNi$_2$B$_2$C}

\author{P. Starowicz}
\affiliation{Ames Laboratory and Department of Physics and
Astronomy, Iowa State University, Ames, IA 50011, USA}
\affiliation{M. Smoluchowski Institute of Physics, Jagiellonian
University, Reymonta 4, 30-059 Krak\'{o}w, Poland}

\author{C.~Liu}
\affiliation{Ames Laboratory and Department of Physics and Astronomy, Iowa State University, Ames, IA 50011, USA}

\author{R.~Khasanov}
\affiliation{Ames Laboratory and Department of Physics and Astronomy, Iowa State University, Ames, IA 50011, USA}
\affiliation{Physik-Institut der Universit\"{a}t Z\"{u}rich,
Winterthurerstrasse 190, CH-8057 Z\"urich, Switzerland}

\author{T.~Kondo}
\affiliation{Ames Laboratory and Department of Physics and Astronomy, Iowa State University, Ames, IA 50011, USA}

\author{G.~Samolyuk}
\affiliation{Ames Laboratory and Department of Physics and Astronomy, Iowa State University, Ames, IA 50011, USA}

\author{D.~Gardenghi}
\affiliation{Ames Laboratory and Department of Physics and Astronomy, Iowa State University, Ames, IA 50011, USA}
\affiliation{Bob Jones University, Greenville, SC 29614, USA}

\author{Y.~Lee}
\affiliation{Ames Laboratory and Department of Physics and Astronomy, Iowa State University, Ames, IA 50011, USA}

\author{T. Ohta}
\affiliation{Advanced Light Source, Berkeley National Laboratory, Berkeley, CA 94720, USA}

\author{B. Harmon}
\affiliation{Ames Laboratory and Department of Physics and Astronomy, Iowa State University, Ames, IA 50011, USA}

\author{P. Canfield}
\affiliation{Ames Laboratory and Department of Physics and Astronomy, Iowa State University, Ames, IA 50011, USA}

\author{S. Bud'ko}
\affiliation{Ames Laboratory and Department of Physics and Astronomy, Iowa State University, Ames, IA 50011, USA}

\author{E. Rotenberg}
\affiliation{Advanced Light Source, Berkeley National Laboratory, Berkeley, CA 94720, USA}

\author{A. Kaminski}
\affiliation{Ames Laboratory and Department of Physics and Astronomy, Iowa State University, Ames, IA 50011, USA}

\date{\today}
\begin{abstract}
We measured the Fermi surface (FS), band dispersion and superconducting gap in LuNi$_2$B$_2$C using Angle Resolved
Photoemission Spectroscopy. Experimental data were compared with the tight-binding version of the Linear Muffin-Tin Orbital (LMTO)
method and Linearized Augmented Plane-Wave (LAPW) calculations. We found reasonable agreement between the two calculations and experimental data. The measured FS exhibits large parallel regions with a nesting vector that agrees with a previous positron annihilation study and calculations of the generalized susceptibility. The measured dispersion curves also agree reasonably well with the TB-LMTO calculations, albeit with some differences in the strength of the hybridization. In addition, the spectrum in the superconducting state revealed a 2meV superconducting gap. The data also clearly shows the presence of a coherent peak above the chemical potential, $\mu$ that  originates from thermally excited electrons above the energy of 2$\Delta$. This feature was not previously observed in the Lu-based material.
\end{abstract}

\pacs{74.70.Dd, 71.18.+y, 71.20.-b, 71.27.+a}

\maketitle

\section{Introduction}
\begin{figure*}
\includegraphics[width=6.25in]{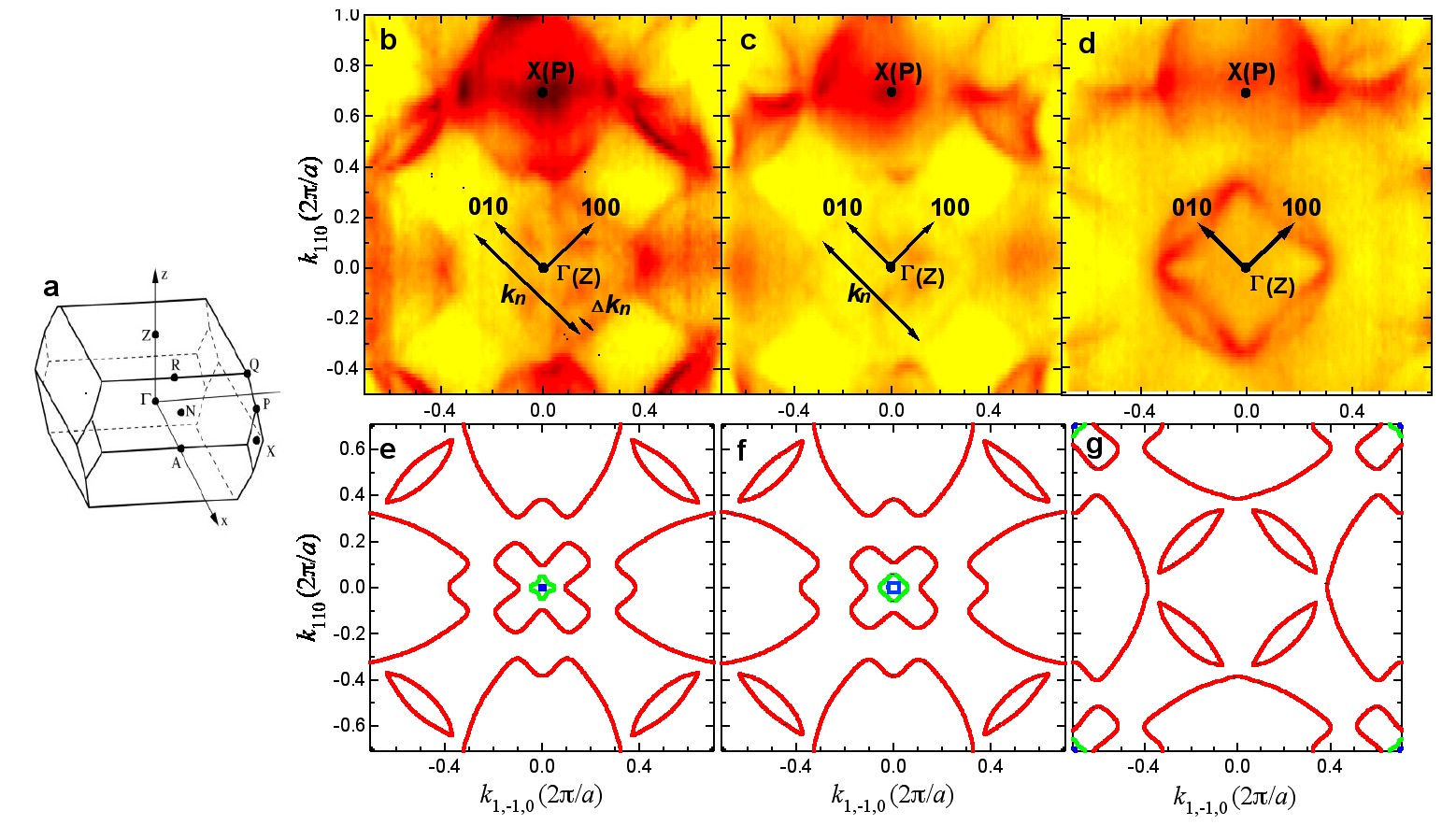}
\caption{(Color online) Comparison between the Fermi surface maps measured by ARPES and the linear muffin-tin orbital (TB-LMTO)
calculation.
(a) Sketch of the first Brillouin zone for LuNi$_2$B$_2$C.
(b-d) ARPES mapping at the chemical potential
for incident photon energies of 128.13 eV, 119.44 eV, and 102.98 eV, respectively.
(e-g) The Fermi surface maps obtained by TB-LMTO
calculations for constant $k_{z}$ values equal to (e) 0.2, (f) 0.15 and (g) 0.8 expressed in the units of $\Gamma$ - Z distance. } \label{fig1}
\end{figure*}

Rare earth nickel borocarbides RNi$_2$B$_2$C (R - rare earth)
constitute an interesting class of materials
\cite{Cava1994,Nagarajan1994,Canfield1998,MullerNarozhnyi2001,MazumdarNagarajan2005},
in which there is a competition and coexistence between
superconductivity and magnetism. Amongst these compounds,
nonmagnetic LuNi$_2$B$_2$C has the highest superconducting critical
temperature of 16.6 K \cite{Cava1994}. The borocarbides exhibit a
peculiar anisotropy of the superconducting gap, the character of
which is still under debate. It is believed that the gap is highly
anisotropic in the two non-magnetic compounds LuNi$_2$B$_2$C and
YNi$_2$B$_2$C
\cite{Boaknin2001,Bobrov2005,Maki2002,Raychaudhuri2004,
MartinezSamper2003,Izawa2002,Yokoya2000}. Its symmetry was proposed
to be \textit{s} + \textit{g} \cite{Maki2002}, which is consistent
with certain experimental results \cite{Raychaudhuri2004} but an
anisotropic s-wave symmetry has also been considered
\cite{MartinezSamper2003}. Other experimental data indicate that the
gap in YNi$_2$B$_2$C has point nodes along the (100) and (010)
directions \cite{Izawa2002}.  LuNi$_2$B$_2$C crystallizes in a
body-centered tetragonal structure with lattice parameters
\textit{a} = 3.4639 \AA , and \textit{c} = 10.6313 \AA
\cite{Siegrist1994}. Its crystal structure consists of  Lu-C layers
with Ni$_2$B$_2$ sheets in between. Previously calculations reveal
that LuNi$_2$B$_2$C is characterised by a large density of states
(DOS) at the Fermi energy ($E_{F}$) originating mainly, but not
exclusively, from Ni \textit{d} electrons
\cite{Mattheiss1994,PickettSingh1994,Coehoorn1994}. Another
interesting feature is a flat band along the $\Gamma$-X direction
just above $E_{F}$. The Fermi surface (FS) topography of
LuNi$_2$B$_2$C was studied by ab-initio calculations
\cite{Rhee1995,Kim1995,Dugdale1999}. Band structure calculations
\cite{Rhee1995} revealed a pronounced maximum in the generalized
electronic susceptibility at ($\sim$0.6\textit{a}*, 0, 0), where
\textit{a}* $\equiv 2\pi/a$ and most likely arising from large
nested regions of the FS. Moreover, phonon softening was observed in
LuNi$_2$B$_2$C by means of inelastic neutron scattering for a range
of wave vectors around (0.5\textit{a}*, 0, 0) \cite{Dervenagas1995}.
Interestingly enough the magnetic ordering, which was found in
RNi$_2$B$_2$C compounds with magnetic atoms R = Er, Ho, Tb and Gd
manifest similar modulation vector usually close to
(0.55\textit{a}*,0,0)\cite{MullerNarozhnyi2001}. The first
experimental studies of the LuNi$_2$B$_2$C (RNi$_2$B$_2$C) Fermi
surface were performed by means of two-dimensional angular
correlation of electron-positron annihilation radiation (2D-ACAR)
and the data were compared to the Linear Muffin-Tin Orbital (LMTO),
local density approximation (LDA) calculations \cite{Dugdale1999}.
Nested parts of the FS were found with a nesting vector
corresponding to both the phonon softening and the magnetic
modulation vectors. The fraction of the FS participating in nesting
was determined to be 4.4 $\pm$ 0.5\% \cite{Dugdale1999}. That study
was however limited only to a rough ``calipering" of the Fermi
surface. Knowledge of the experimental band structure, Fermi surface
and quasiparticle properties is deemed essential to understand the
interplay of the various interactions in these materials, as it may
shed new light on other phenomena such as anisotropic
superconductivity, the role of phonon softening and the relationship
between the superconductivity and magnetic ordering in the
borocarbides. It is also  a pre-requisit for direct determination of
the alleged anisotropy of the superconducting gap in these
materials. In this report we present angle resolved photoelectron
spectroscopy (ARPES) measurements of the band dispersion, Fermi
surface and supercondcuting gap in the borocarbide with the highest
$T_{c}$, LuNi$_2$B$_2$C. The experimental results were compared with
the tight-bounding LMTO (TB-LMTO) method and the full potential LAPW calculations. 
We found reasonable agreement with
theory. The most significant difference between the calculations and
experimental data is the strength of the hybridization. We also
determined the superconducting gap  to be 2.58 meV (extrapolated for T=0), in
good agreement with the gap expected from the superconducting
transition temperature (${{2\Delta}/{k_BT_C}}=2.78$).

\section{Experimental}

LuNi$_2$B$_2$C single crystals were grown at Ames Laboratory by
means of a high-temperature flux
technique\cite{Canfield1998,Fisher2001}. The plate-like crystals
were cleaved \textit{in situ} at pressures better than 3 x
10$^{-11}$ Tr to reveal and maintain fresh \textit{a}-\textit{b}
surfaces. The Fermi surface and band structure mapping were
performed at the 7.0.1 beamline at the Advanced Light Source, using
a Scienta R4000 analyzer.  The energy and angle resolution were set
at $\sim$ 30 meV and $\sim$ 0.5 deg, respectively. The energy gap
was measured with a Scienta 2002 analyser and He-I photon source
(\textit{h}$\nu$ = 21.2 eV), in which the overall energy resolution
was set at 2 meV. The normal state data were measured at the
Synchrotron Radiation Center using the PGM beamline and Scienta 2002
endstation, with the energy and angular resolution set at $\sim$ 13
meV and 0.25 deg, respectively. Tight-binding linear muffin-tin
orbital calculations were performed by the TB-LMTO program, version
47 \cite{LMTO}, and the Full-Potential Linearized Augmented Plane-Wave
(LAPW) calculations were performed using the Wien2k package \cite{Wien}.

\begin{figure*}
\includegraphics[width=7in]{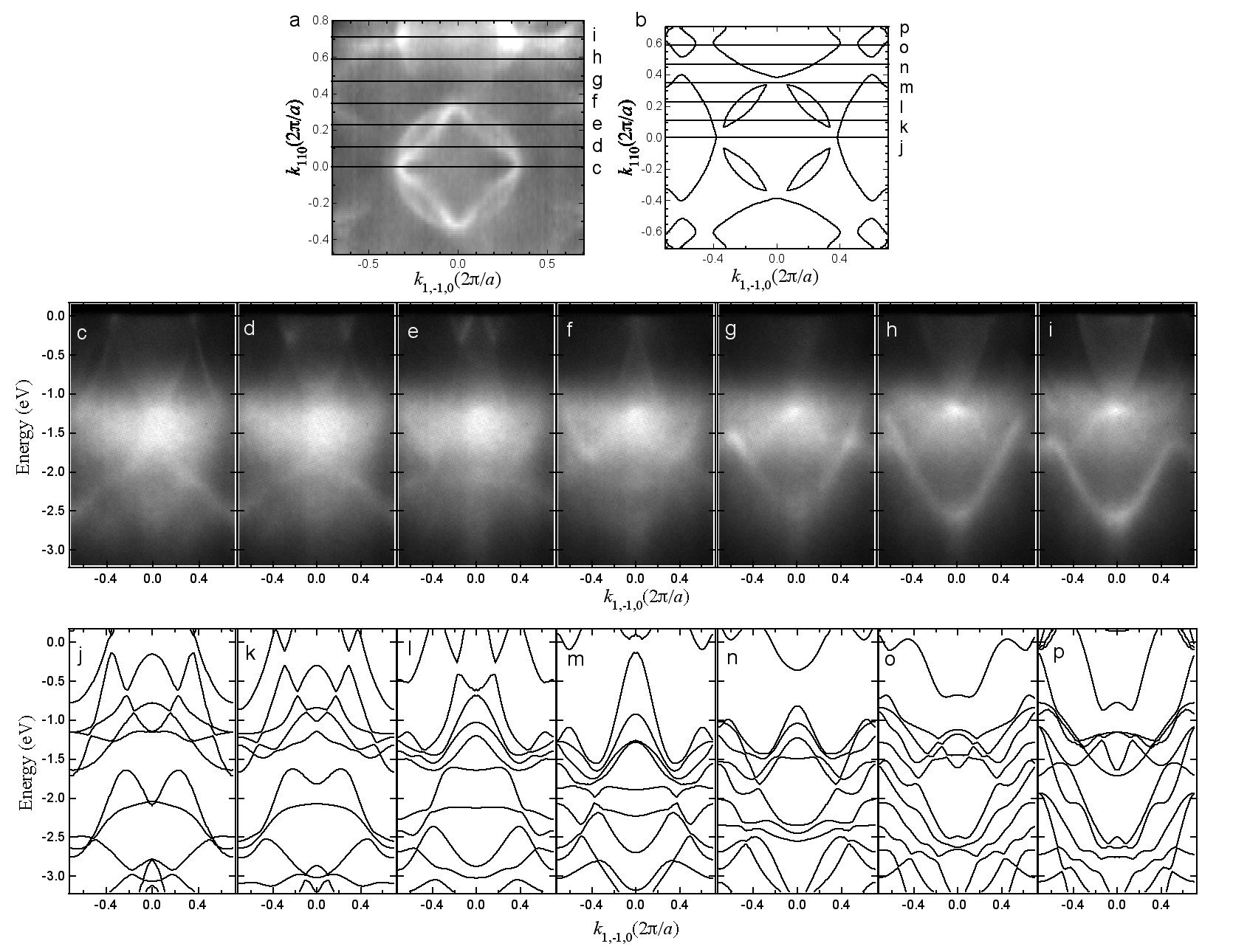}
\caption{Dispersion of the conduction bands obtained
with a photon energy of 102.98 eV (same as Fig. 1d), compared with
the TB-LMTO calculation. a) Fermi surface map with momentum cuts
indicated by the solid lines, in which cut (c) and (i) pass through
the $\Gamma$ and X points, respectively. b) Fermi surface contours
obtained by the TB-LMTO calculation for the value of $k_z$
corresponding to the data in panel (a). Panels (c-i): measured band
dispersion along the cuts indicated in (a). Panels (j-p): calculated
band dispersion along the cuts marked in panel (b).} \label{fig2}
\end{figure*}

\section{Results and discussion}

Band structure and semi-planar Fermi Surface cuts were determined
for incident photon energies 128.13 eV, 119.44 eV and 102.98 eV
(Fig. 1b-d), where the $\Gamma$ point in the Brillouin zone (Fig.
1a) corresponds to normal emission of electrons along the (001)
direction. A $\sin r/r$ correction term ($r$ is the distance from
the $\Gamma$ point) was used to account for mapping of the momentum
space onto the angular distribution of photoelectrons. The ARPES
process in 3D materials leaves some ambiguity as to the $k_{z}$
component of the momentum (perpendicular to the sample surface),
because it is not conserved in the photoemission process. This is
due to jump of the potential at the sample surface. From
conservation of energy and remaining components of the momentum one
can calculate the relative changes of the $k_{z}$ for various photon
energies. To obtain the offset one needs to seek guidance from the
band structure calculations and identify the high symmetry points in
the data.\cite{HUFNER} This allows estimation of the $k_{z}$
offset. The change of the wave vector component $k_{z}$ (parallel
to the \textit{c} axis) between scans in Fig. 1b ($h\nu$=128.13) and
1c ($h\nu$=119.44) was calculated from momentum and energy
conservation to be 0.33 of the $\Gamma$-Z distance. Similarly the
change of the wave vector component $k_{z}$ between scans in  Fig.
1c ($h\nu$=119.44) and 1d ($h\nu$=102.98) was 0.66 of the
$\Gamma$-Z. The calculated Fermi surfaces were obtained for constant
$k_{z}$ values by means of the TB-LMTO method and are shown in Fig.
1 panels e-g. We estimated the values of the inner potential,
$V_{0}$ = 9.4 eV and the work function $\phi$ = 4.6 eV, by comparing
the high symmetry points between the calculated Fermi surfaces and
the experimental data. This allowed us to determine the offset of
the photon energy that corresponds to $k_{z}$ = 0.

The Fermi surface maps for the incident photon energies of 128.13 eV
and 119.44 eV reveal large parallel parts of the FS with essentially the
same nesting vector (spacing between the linear sections): $k_{n}$ =
(0.59 $\pm$ 0.04)\textit{a}* for Fig. 1b and $k_{n}$ = (0.58 $\pm$
0.04)\textit{a}* for Fig. 1c. $k_{z}$ is expressed in the units of
the $\Gamma$-Z distance, where the $\Gamma$ point corresponds to
$k_{z}$ = 0. Although the full three dimensional FS was not
determined in great detail, a similar nesting vector was found for
different $k_{z}$ values which indicates that the FS likely has
considerable nesting properties for a wide range of $k_{z}$ values.
The constancy of the value of the nesting vector between $k_{z}$=0.15
and -0.2 is also consistent with results of calculations. The
spacing between the parallel segments of the Fermi surface predicted
by TB-LMTO calculation is between 0.54\textit{a}* and
0.55\textit{a}*, the LAPW calculation results (not shown) are
0.50\textit{a}* and 0.57\textit{a}*, respectively. The detected
$k_{n}$ is very close to the theoretically predicted value obtained
from the generalized susceptibility \cite{Rhee1995}. Our results
also agree reasonably well with the nesting vector previously
determined via 2D-ACAR \cite{Dugdale1999}.

The Fermi surface map obtained at 102.98 eV very closely resembles
the calculated Fermi surface for $k_{z}$ = 0.8\textit{a}*. The
overall shapes of the measured and calculated Fermi surface sheets
(Fig. 1d and 1g) are very similar, however there is one significant
difference. In the calculations the four oval parts of the Fermi
surface centered about $\Gamma$-Z are well separated in momentum
space (Fig. 1g), while the data reveals
that they actually are connected at the edges (Fig. 1d). A lack of
separation in the experimental data may indicate that the
hybridization gap is overestimated in the calculations. These oval
parts arise from the intersection of the electron and hole-like
bands. Interestingly enough at the edges along the diagonal
directions (e. g. 110) the bottom of the electron band and the top of
the hole band appear to be pinned at the chemical potential,
resulting in a characteristic ``flower" shape.

In Fig. 2 we plot the band dispersion data along a few selected cuts
in momentum space obtained at an incident photon energy of 102.98 eV
(Figs. 2c-i), along with a calculated (TB-LMTO method) band
dispersion for $k_{z} \sim$ 0.8\textit{a}* (Figs. 2j-p). The
agreement between the measured and calculated band dispersion is
rather good, especially in the proximity of the chemical potential. 
In the corresponding TB-LMTO calculations (Figs. 2j-p),
the same overall features are well reproduced, which shows the
validity of the calculation in this material to a certain extent.
This agreement also validates the assignment of $k_{z}$ values to
the cuts measured at various photon energies, which is very
important when studying 3D materials with ARPES. The most
significant difference is the hybridization gap, which is quite
large in the calculations but its signatures are for the most part
absent in the measured data. For example in Fig. 2 panels (k) and
(l) the high and low energy branches form hybridization gap of about
200 meV at \textit{E} = -0.3 eV, while in the corresponding measured
data (panels (d) and (e)) the bands appear to disperse without a
signature of the hybridization gap. One should consider if the
observed features have any relation to the superconducting gap
asymmetry and observed phonon softening in LuNi$_2$B$_2$C. It should
be noted that band structure calculations
\cite{PickettSingh1994,Kim1995} show a flat band lies very close
to, but slightly above, the Fermi level. This feature was
unfortunately not observed in our data due to the Fermi function
cut-off. However, a higher DOS near the Fermi level would explain
the large number of scattered electrons observed with \textit{k}
vectors along (110) and phonon softening for the discussed wave
vectors. Consequently this may lead to an anisotropy of the
superconducting order parameter. This is in agreement with the
results proposing for YNi$_2$B$_2$C that the superconducting gap is
larger just at (110) and diminishes or even has nodes along the
(100) and (010) directions \cite{Izawa2002}.

Given the above concern, we measured the energy gap in
LuNi$_2$B$_2$C by partial angle-integrated photoelectron
spectroscopy and compared with the normal state Fermi surface. The
opening of the superconducting gap is clearly shown in Fig. 3. In
order to determine the magnitude of the gap, the Dynes function
\cite{Dynes1978} was fitted to the symmetrized \cite{Norman1998}
spectrum (Fig. 3b). The fitted function yields the gap value of
$\Delta$ = 1.5 meV for the sample at \textit{T} = 11 K with the
$\Gamma$ parameter equal to 0.05 meV.

The striking feature in Fig. 3c is the pronounced peak above the
chemical potential. This peak arises from thermal excitation of
electrons above the 2$\Delta$. This points to high DOS just above
$\mu$ which is consistent with the idea that the flat band along
(110) direction, a large part of which is slightly over the Fermi
energy playing an important role in this anisotropic
superconductivity. Similar peaks were recently reported in Y based
borocarbides \cite{YOKOYA2007}. According to BCS theory the energy gap value at zero temperature ($\Delta_0$) is
2.45 meV for $T_{c}$ = 16 K superconductor. Our $\Delta$ value of 1.5 meV obtained at T=11K corresponds to ($\Delta_0$)=2.6 meV, 
in excellent agreement with the BCS predictions.

\begin{figure}
\includegraphics[width=3.25in]{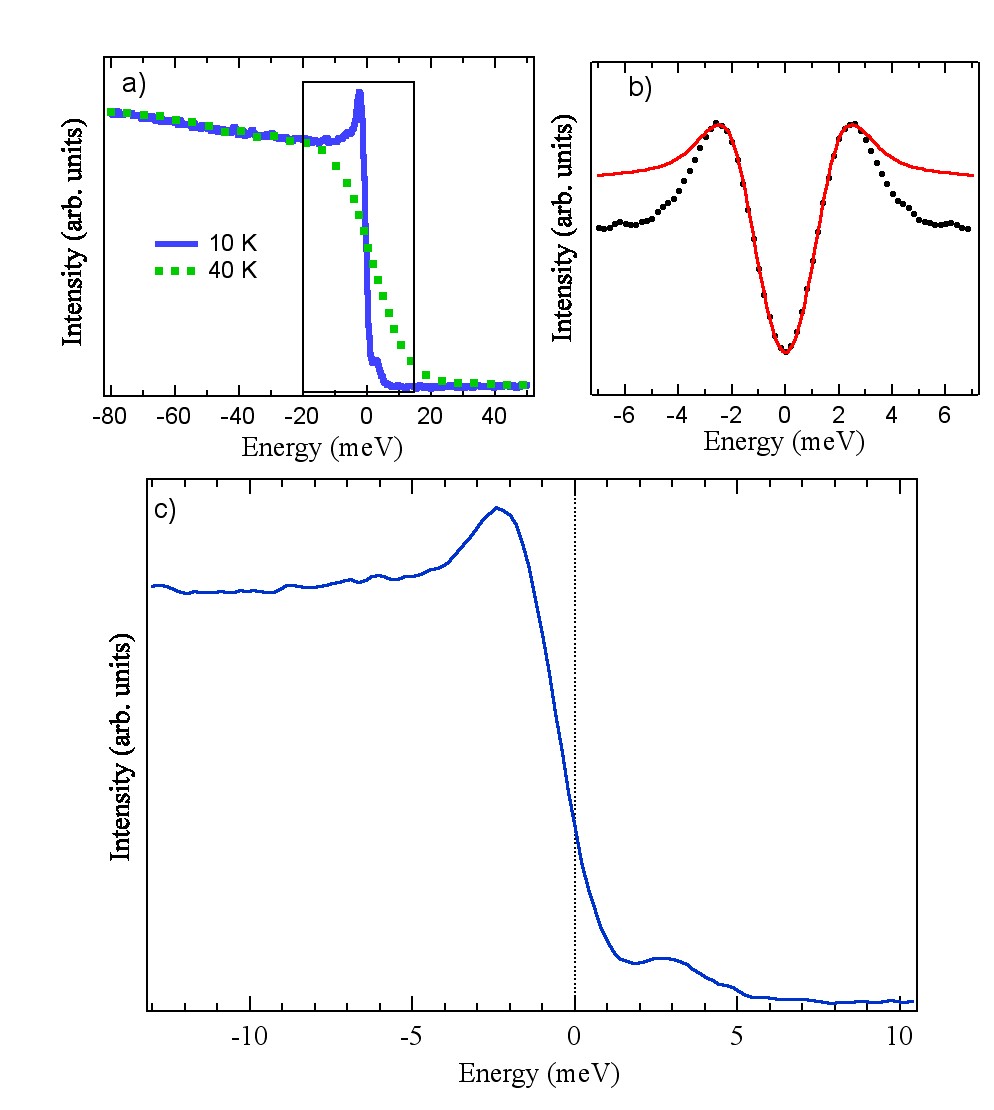}
\caption{(Color online) Superconducting gap of LuNi$_2$B$_2$C a)
measured at \textit{T} = 11$\pm$1 K, compared with the normal state at
\textit{T} = 40 K. b) The Dynes function (solid red line) with the
parameters $\Delta$ = 1.5 meV and $\Gamma$ = 0.05 meV fitted to the
symmetrised spectrum (solid black circles). c) enlarged portion of
superconducting spectra from (a) close to the chemical potential.}
\label{fig3}
\end{figure}

\section{Conclusions}

We have performed measurements of the Fermi surface, band dispersion
and supercondcuting gap for highest $T_c$ rare earth nickel borocarbide
superconductor LuNi$_2$B$_2$C. The experimental data were compared
with two different density functional calculations. The overall agreement between theory and measurement
is good. In the experiment, large parallel FS parts spaced with the
vector $k_{n}$ = 0.59\textit{a}* have been found for two different
incident photon energies, which is a confirmation of the previous
theoretical predictions \cite{Rhee1995} and earlier experimental
studies \cite{Dugdale1999}. The calculated FS confirms the existence
of large nested parts, with a nesting vector in good agreement with
the ARPES results presented here. 
The superconducting gap was measured and we also observed a coherent
peak above the chemical potential. This peak arises due to electrons
being thermally excited above the energy of 2$\Delta$.

\section{Acknowledgments}

This work was supported by Director Office for Basic Energy
Sciences, US DOE. Work at Ames Laboratory was supported by the
Department of Energy - Basic Energy Sciences under Contract
No. DE-AC02-07CH11358. Advanced Light Source is operated by the U.S. DOE under
Contract No. DE-AC03-76SF00098. Synchrotron Radiation Center is
supported by the  National Science Foundation under award No. DMR-0537588.
R. K. gratefully acknowledges support of K. Alex M\"uller Foundation.

\end{document}